\documentclass[letterpaper]{article} % DO NOT CHANGE THIS
\usepackage[preprint]{aaai2027} % Non-anonymous arXiv preprint; no AAAI copyright notice.
\usepackage[hyphens]{url}  % DO NOT CHANGE THIS
\usepackage{graphicx} % DO NOT CHANGE THIS
\usepackage{natbib}  % DO NOT CHANGE THIS AND DO NOT ADD ANY OPTIONS TO IT
\usepackage{caption} % DO NOT CHANGE THIS AND DO NOT ADD ANY OPTIONS TO IT
\usepackage{algorithm}
\usepackage{algorithmic}
\usepackage{listings}
\usepackage[table]{xcolor}
\usepackage{booktabs}
\usepackage{tabularx}
\usepackage{array}

\lstdefinestyle{casestudydiff}{
    basicstyle=\ttfamily\scriptsize,
    numbers=none,
    breaklines=true,
    breakatwhitespace=false,
    columns=fullflexible,
    keepspaces=true,
    showstringspaces=false,
    frame=single,
    rulecolor=\color{black!35},
    backgroundcolor=\color{black!2},
    xleftmargin=0.3em,
    xrightmargin=0.3em,
    aboveskip=0.4em,
    belowskip=0.4em
}
\usepackage{newfloat}
\DeclareCaptionStyle{ruled}{labelfont=normalfont,labelsep=colon,strut=off} % DO NOT CHANGE THIS
\floatstyle{ruled}
\newfloat{listing}{tb}{lst}{}
\floatname{listing}{Listing}

\usepackage{multirow}
\usepackage{amsmath,amssymb}
\usepackage{threeparttable}

\definecolor{scopefg}{RGB}{41,98,186}
\definecolor{subjectfg}{RGB}{22,163,74}
\definecolor{typefg}{RGB}{180,83,9}

\newcommand{\typeword}[1]{\textcolor{typefg}{#1}}
\newcommand{\scopeword}[1]{\textcolor{scopefg}{#1}}
\newcommand{\subjectword}[1]{\textcolor{subjectfg}{#1}}

\title{CoCoRerank: Towards Conventional Commit Message Generation by \\Component and Candidate Consistency Reranking}
\author{
    Shaopeng Jia\equalcontrib,
    Yali Du\equalcontrib,
    Ming Li\corresponding
}
\affiliations{
    National Key Laboratory for Novel Software Technology,\\
    School of Artificial Intelligence, Nanjing University, China\\
    \{jiasp, duyl, lim\}@lamda.nju.edu.cn
}

\begin{document}

\maketitle

\begin{abstract}
Commit messages are essential for understanding software changes, yet
automatic commit message generation typically treats a message as an
unstructured text sequence. This limits its ability to support
standardized development workflows, where commit messages are often
expected to follow the Conventional Commits Specification (CCS) in the
form \typeword{type} (\scopeword{scope}): \subjectword{subject}. In this paper, we study conventional
commit message generation under the complete CCS format. We construct a new benchmark of 86,688 high-quality commits collected from
open-source GitHub repositories, with each message normalized into
\typeword{type}, \scopeword{scope}, and \subjectword{subject} through structural
normalization and semantic quality filtering. Based on this benchmark, we
propose a two-dimensional consistency-based reranking framework named \textsc{CoCoRerank} for
LLM-based generation. \textsc{CoCoRerank} exploits horizontal consistency
among the code change, type, scope, and subject, as well as vertical
consensus across multiple generated candidates. Experiments with
representative CMG baselines, LLM generators, reranking strategies, and
ablation variants show that \textsc{CoCoRerank} improves both structural
component prediction and subject generation quality. The results demonstrate
that complete CCS supervision and multidimensional consistency modeling
provide an effective foundation for accurate and standardized commit
message generation. The artifact is publicly released at
\url{https://github.com/bluewhalebug/CoCoRerank}.
\end{abstract}

\section{Introduction}
\label{sec:introduction}
A commit message records the intent and effect of a software change in
natural language. It is routinely used in code review, debugging,
maintenance, release management, and collaboration, where developers
need to understand not only what code was changed but also why the
change matters \cite{Bar2016relation,Li2023matter}. However, writing an informative commit message is a
manual and often neglected activity. This tension has motivated commit
message generation (CMG), which aims to generate a message directly from
a code diff. 
Recently, both task-specific learning methods and large language
models (LLMs) have been widely applied to software engineering tasks,
such as code
summarization~\cite{crupi2025effectiveness,virk2025calibration}, code generation~\cite{sun2026aces,liurandom,du2026design,du2026cit}, bug localization~\cite{du2023pre,ma2023capturing}, and code translation~\cite{du2023beyond,du2024joint,du2025post}. In CMG, these models typically operate on code diffs,
which provide a compact representation of the changes introduced by a
commit rather than the complete source code. 
Early neural approaches formulated CMG as a
seq-to-seq translation problem \cite{jiang2017automatically}, and later
methods improved generation by incorporating code structure, edit
representations, retrieval signals, pretraining, and hybrid
generation--retrieval designs
\cite{xu2019commit,nie2021coregen,dong2022fira,liu2018neural,liu2020atom,shi2022race,he2023come}.
These advances have improved the fluency and informativeness of
generated messages, but they mostly treat a commit message as a single
unstructured text sequence.

Unstructured generation is insufficient for practical development workflows as real projects often rely on standardized commit formats. The \textbf{Conventional Commits Specification (CCS)} represents a
commit message as \typeword{type} (\scopeword{scope}): \subjectword{subject}, where \typeword{type}
indicates the category of a change, \scopeword{scope} identifies the
affected component, and \subjectword{subject} summarizes the concrete
modification. This decomposition makes commit messages both human
readable and machine interpretable, enabling downstream automation such
as changelog generation and semantic versioning. It also exposes a more
fine-grained generation objective: a conventional message is correct only
when its category, affected component, and textual description are
mutually compatible with the same code change. Existing CMG methods do
not explicitly model this three-part structure. A generated message
may be fluent while still assigning an inconsistent type, an irrelevant
scope, or a subject that does not match the predicted metadata.

Progress on conventional commit message generation is further limited by
the lack of suitable datasets \cite{Zeng2025look}. Existing CMG datasets have supported
important progress in neural generation, project-specific modeling, and
large-scale evaluation
\cite{jiang2017automatically,liu2019generating,xu2019commit,tao2021evaluation,schall2024commitbench}.
Nevertheless, they generally provide unstructured messages or retain
only partial structured information, which prevents models from being
trained and evaluated on the complete CCS format. As a result, current
evaluation cannot separately measure whether a model predicts the right
change type, identifies the right scope, and writes a subject that is
consistent with both. This gap makes it difficult to study the
central question: how to generate conventional commit messages whose
components are individually accurate and jointly coherent.

To address this gap, we construct a new dataset for conventional commit
message generation. The dataset contains 86,688 high-quality commits
collected from open-source GitHub repositories, and each message is
normalized into the complete \typeword{type}, \scopeword{scope}, and
\subjectword{subject} structure. We apply systematic structural
normalization to remove malformed or ambiguous CCS fields and semantic
quality filtering to discard vague, automatically generated, or
diff-inconsistent messages. By preserving the complete structure, the
dataset provides supervision for conventional generation and enables
fine-grained evaluation of each component.

Built upon the dataset, we present \textsc{CoCoRerank}, a dual-consistency reranking framework for LLM-driven conventional commit message generation. Its name derives from the Conventional Commits Specification (CCS) and the two core consistency constraints: component and candidate consistency. Modern LLMs readily generate multiple candidate commits from a single code diff, yet optimal candidate selection remains difficult \cite{li2022competition, chen2023codet, liu2026dynamic}. This problem becomes more severe for conventional CMG, as candidates differ widely in structural compliance and semantic alignment with code edits. Our core insight is that high-quality CCS commits must satisfy two complementary consistency requirements: 1) \textbf{Horizontal component consistency}: The \typeword{type}, \scopeword{scope}, and \subjectword{subject} segments act as mutually complementary semantic views and must coherently describe the same code change; prior work confirms cross-view consensus boosts output reliability \cite{huang2024enhancing,sun2023enhancing}. 2) \textbf{Vertical candidate consistency}: Valid change interpretations recur across multiple generated candidates, a self-consistency property proven effective for filtering trustworthy model outputs \cite{wang2023selfconsistency, chen2024universal}. In \textsc{CoCoRerank}, we extract consensus type and scope from all candidates, score each subject’s compatibility with the code diff and consensus metadata, and aggregate scores over semantically clustered subjects. This pipeline picks commits that are both structurally coherent and statistically representative of the full candidate pool.

We evaluate the proposed dataset and \textsc{CoCoRerank} against
representative retrieval-based, learning-based, hybrid, and advanced LLM-based
baselines. Experiments show that our method improves direct LLM
generation across the conventional components and outperforms alternative
candidate selection strategies such as majority voting and
minimum-Bayes-risk reranking. Ablation and cross-model analyses further
show that both component consistency and candidate consensus contribute
to the final performance.
Our main contributions are summarized as follows:
\begin{itemize}
    \item We introduce a new dataset containing 86,688 high-quality
    commits that strictly follow the CCS format. Unlike existing CMG
    datasets, it preserves the complete \typeword{type}, \scopeword{scope},
    and \subjectword{subject} structure, constructed through rigorous structural and semantic quality control.

    \item We propose a two-dimensional
    consistency-based reranking method named \textsc{CoCoRerank} for LLM-based commit message
    generation. The method jointly
    exploits consistency among the three CCS components and consistency
    across multiple generated candidates, leading to more accurate and
    structurally coherent commit messages.

    \item We conduct extensive experiments with representative CMG
    baselines, LLM generators, and reranking strategies. The results demonstrate the effectiveness of
    CCS modeling and show that the proposed consistency signals are
    complementary to conventioncal CMG.
\end{itemize}

\section{Related Work}
\label{sec:related_work}

In this section, we review related work on commit message generation from two perspectives: methods and datasets.

\subsection{Commit Message Generation Methods}
\label{sec:related_methods}

Existing commit message generation methods can be broadly divided into rule-based, retrieval-based, learning-based, and hybrid methods.
\textbf{Rule-based} methods summarize code changes using predefined rules or templates \cite{buse2010automatically,cortescoy2014changescribe,shen2016automatic}. However, their reliance on handcrafted rules and language-specific analysis limits generality.
\textbf{Retrieval-based} methods generate messages by reusing those of similar historical commits. NNGen \cite{liu2018neural} retrieves candidates using bag-of-words similarity and reranks them by diff similarity, while LogGen \cite{hoang2020cc2vec} retrieves similar patches using CC2Vec representations. Such methods may fail to adapt retrieved messages to new changes and can introduce irrelevant details.
\textbf{Learning-based} methods directly learn the mapping from code changes to natural-language descriptions. 
For instance, NMT \cite{jiang2017automatically} formulates CMG as neural machine translation with an attentional RNN encoder--decoder. CODISUM \cite{xu2019commit} incorporates code structure and a copy mechanism, CoreGen \cite{nie2021coregen} pretrains a Transformer on code changes, and FIRA \cite{dong2022fira} models changes as fine-grained edit graphs. \ensuremath{\textrm{C}^4\textrm{MG}}~\cite{du2025capturing} leverages dynamic control flow graphs to capture context-aware code changes for improving commit message generation. 
\textbf{Hybrid} methods combine retrieval and generation. ATOM \cite{liu2020atom} selects between generated and retrieved messages, RACE \cite{shi2022race} uses retrieved commits as generation exemplars, and COME \cite{he2023come} combines translation and retrieval through a decision model.

%Overall, most of the existing methods treat commit messages as unstructured text, overlooking the explicit organization provided by the Conventional Commits Specification. These limitations motivate us to investigate structured commit message generation by jointly modeling the consistency among the \text{type}, \text{scope}, \text{subject} of the commit message.

\subsection{Commit Message Generation Datasets}
\label{sec:related_datasets}

Several datasets have been built to advance research on commit message generation.
CommitGenData \cite{jiang2017automatically} and CoDiSumData \cite{xu2019commit} are both sourced from GitHub’s top 1,000 Java repositories.
PtrGNCMsgData \cite{liu2019generating} aggregates commits from around 2,000 Java repositories and was originally proposed to test pointer-generator models against project-specific identifiers and out-of-vocabulary tokens.
MCMD \cite{tao2021evaluation} covers roughly 2.25 million commits across 500 repositories, equipped with abundant metadata and diverse data partition schemes. That said, its lenient filtering criteria inevitably lead to redundant overlapping entries and noisy samples.
By contrast, CommitBench \cite{schall2024commitbench} collects 1,664,590 commits from 71,676 projects, with thorough filtering to eliminate noise, duplicates, privacy-sensitive records, and other irrelevant content.
CommitSuite \cite{wan2026commitsuitecomprehensivebenchmarkcommit} consists of 63,533 commits compliant with the CCS format mined from 243 repositories, featuring AST-level code diff information and LLM-aided annotations for commit \emph{what} and \emph{why} semantics. However, its generation setup only reserves the commit type and plain text description, completely stripping out the scope field.

\begin{table*}[h]
\small
\centering
        \begin{tabular}{lcccccccc}
        \toprule
         \multirow{2}{*}{\textbf{Dataset}} & \multicolumn{4}{c}{\textbf{Commit Message}} & \multicolumn{3}{c}{\textbf{Commit}} & \multirow{2}{*}{\textbf{Size}} \\ \cmidrule(lr){2-5}\cmidrule(lr){6-8} 
          & \textbf{Type} & \textbf{Scope} &\textbf{Subject} & \textbf{Subject Length} & \textbf{Code Diff} & \textbf{Context} & \textbf{Diff Length} & \\
        \midrule
        CommitGen~\shortcite{jiang2017automatically} & -- & -- & V-DO & $\leq 30$ tokens & \checkmark & -- & $\leq 100$ tokens & 32,208 \\
        PtrGNCMsg~\shortcite{liu2019generating} & -- & -- & V-DO  & $\leq 30$ tokens & \checkmark & -- & $\leq 100$ tokens & 32,663 \\
        CoDiSum~\shortcite{xu2019commit} & -- & -- & -- & $\leq 20$ tokens & \checkmark & -- & $\leq 200$ tokens & 90,661 \\
        MCMD~\shortcite{tao2021evaluation} & -- & -- & -- &  -- & \checkmark & \checkmark & --  & 2,250,000  \\
        CommitBench~\shortcite{schall2024commitbench} & -- & -- & -- & $8$--$512$ tokens & \checkmark & \checkmark & $\leq 512$ tokens & 1,664,590 \\
        CommitSuite~\shortcite{wan2026commitsuitecomprehensivebenchmarkcommit} & \checkmark & -- & -- & IQR-based & \checkmark & \checkmark & IQR-based & 63,533 \\
        \midrule
        \textbf{Ours} & \checkmark & \checkmark & V-DO & $4$--$63$ tokens & \checkmark & \checkmark & $64$--$32767$ tokens & 86,688 \\
        \bottomrule
    \end{tabular}
    \begin{tablenotes}
\footnotesize
\item $^{*}$ "--" indicates that the corresponding feature or restriction is not present or applied in the dataset.
\end{tablenotes}
    \caption{Statistics on restrictive filters in existing CMG datasets.}
    \vspace{0.3cm}
\label{tab:filters3}
    \vspace{-0.5cm}
\end{table*}

\begin{figure*}
    \centering
    \includegraphics[width=\linewidth]{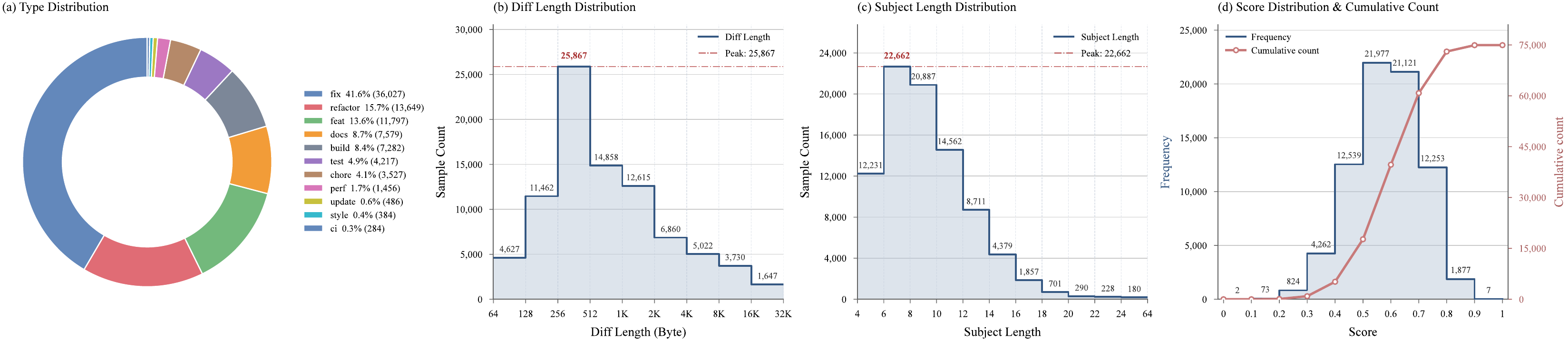}
    \caption{Statistical Distribution Analysis of Sample Type Distribution, Diff Length, Subject Length and Score Distributions}
    \label{fig:placeholder}
\end{figure*}

\section{Dataset}
\begin{figure}[t]
	\centering
	\includegraphics[width=\linewidth]{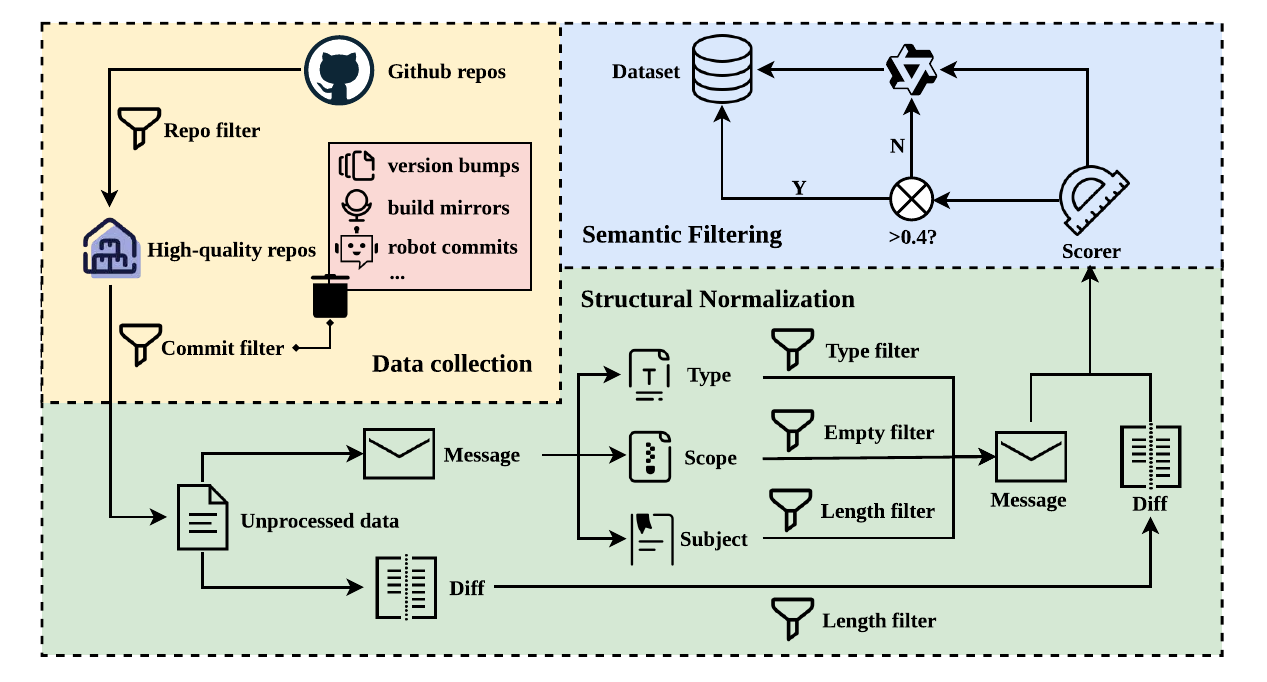}
	\caption{Overview of the dataset construction pipeline.}
	\label{dataset_pipeline}
\end{figure}

%In this section, we briefly outline the construction pipeline of our new CCS-compliant CMG dataset, covering data collection, multi-stage filtering, structural normalization, and semantic quality optimization. We further present the key statistics and dataset partition strategy of the final benchmark. The overall construction workflow is illustrated in Figure \ref{dataset_pipeline}, while fine-grained implementation details, step-by-step filtering metrics, and detailed distribution analyses are provided in the Appendix.
In this section, we briefly summarize the key steps in constructing our
conventional CMG dataset and analyze the final dataset through its key statistics and distributions. Figure~\ref{dataset_pipeline}
illustrates the overall construction workflow, while
fine-grained implementation details are provided in the Appendix.
%Table~\ref{tab:filters3} compares the restrictive filters adopted by existing CMG datasets, and Figure~\ref{fig:placeholder} presents the statistical distributions of the final dataset. 

\subsection{Dataset Construction Overview}

Our dataset construction consists of three stages: \textbf{Data Collection}, \textbf{Structural Normalization}, and \textbf{Semantic Quality Filtering}. In the first stage, we collect commits from 206 GitHub repositories, exclude small repositories with insufficient commit histories, remove meaningless or automatically generated commits, apply verb--direct-object (V-DO) constraints \cite{jiang2017automatically} and token-length filtering to code diffs and commit subjects. Table~\ref{tab:filters3} compares the restrictive filters adopted by existing CMG datasets. In the second stage, we convert all samples into the standard \typeword{type}(\scopeword{scope}): \subjectword{subject} format, remove instances with missing core fields, merge misspelled or fragmented type labels, and clean redundant metadata. In the final stage, we use embedding similarity for initial screening and set 0.4 as the cutoff based on Figure~\ref{fig:placeholder}(d): samples above it are retained directly, while the rest undergo LLM-based quality assessment to remove semantic inconsistencies.

\begin{figure*}[t]
\centering
% Replace the placeholder below with:
\includegraphics[width=\textwidth]{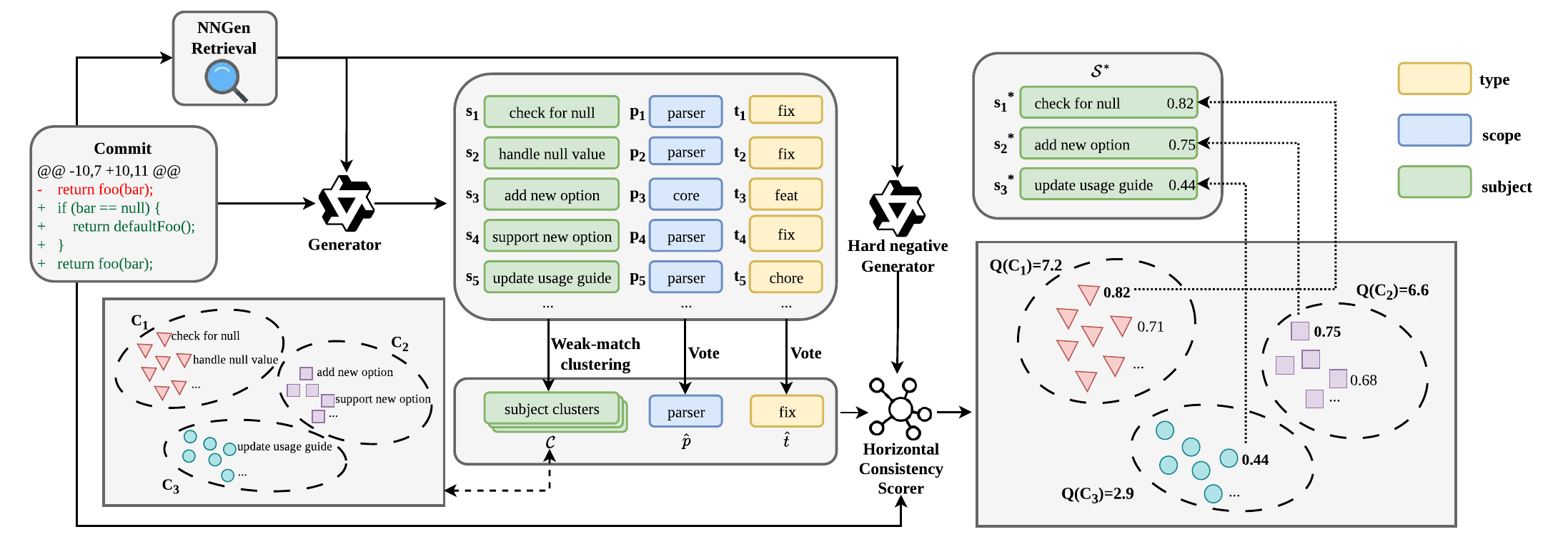} 
\caption{Overview of the \textsc{CoCoRerank} framework.}
\label{fig:method_pipeline}
\end{figure*}

%\subsection{Final Dataset Statistics}
%After multi-stage structural and semantic purification, our final dataset contains 86,688 high-quality CCS-compliant commit-diff pairs. Our dataset covers both fine-grained local edits and large-scale code changes, with concise and standardized commit subjects that conform to real-world development conventions. The natural long-tailed distribution of commit types is preserved to maintain realistic software development characteristics. Finally, we stratify the dataset by commit type and split the samples into training, retrieval, validation, and test sets at a ratio of 6:2:1:1.
 \subsection{Final Dataset Statistics}

 After all structural and semantic filtering steps, the final dataset contains 86,688 high-quality commit--message pairs.

Figure~\ref{fig:placeholder}(b) and~\ref{fig:placeholder}(c) present the token-length distributions of code diffs and commit subjects. All token counts are computed using the Qwen3-0.6B tokenizer. Diff lengths vary widely, with the largest proportion falling in the $[256,512)$ interval. Overall, 65.54\% of diffs contain fewer than 1,024 tokens, while 12.00\% contain at least 4,096 tokens, indicating that the dataset covers both localized and large-scale code changes. Commit subjects are much more concentrated. In total, 96.24\% contain fewer than 16 tokens. Only a small fraction exceed 24 tokens, which conforms to the concise writing habit of industrial commit messages. Figure~\ref{fig:placeholder}(a) shows the distribution of the commit types. The distribution is clearly long-tailed, with \texttt{fix} being the most frequent type, while categories such as \texttt{ci}, \texttt{style}, and \texttt{update} occur much less frequently. We preserve this real-world distribution instead of artificial balancing. The dataset is then stratified by type and divided into training, retrieval, validation, and test sets in a 6:2:1:1 ratio, with near-duplicate commits checked across splits to prevent data leakage.

\section{The Proposed Method: \textsc{CoCoRerank}}
\label{sec:method}
In this section, we present the proposed \textsc{CoCoRerank} framework for conventional commit message generation.
\subsection{Overview} %总体介绍方法包含的组件
\label{sec:method_overview}

Given a code change $x$, our goal is to generate a commit message $m$ that strictly follows the CCS format in the following standardized form: \typeword{type} (\scopeword{scope}): \subjectword{subject}.

  As illustrated in Figure~\ref{fig:method_pipeline}, we propose a two-dimensional consistency-based reranking framework. First, we retrieve similar historical commits as in-context demonstrations and generate multiple conventional candidates. We then obtain consensus predictions for type and scope, evaluate each candidate subject against the commit and the consensus fields, and group weakly matched subjects into clusters. The clusters are ranked by their aggregated consistency scores, and the highest-scoring subject in the top-ranked cluster is selected to construct the final message.

\subsection{Retrieval-Augmented Candidate Generation}
\label{sec:candidate_generation}

For each commit $x$, we retrieve five similar historical commits using the retrieval method of NNGen \cite{liu2018neural}. Similarity is computed between the target diff and historical diffs. Each retrieved example consists of a code diff and its corresponding conventional message. These examples are inserted into the prompt as in-context demonstrations.

Conditioned on the target diff and the retrieved examples, the LLM generates $N=100$ candidate messages  $\mathcal{M}=\{m_i\}_{i=1}^{N}$. Each candidate is constrained to follow the CCS format and is parsed as
\begin{equation}
m_i=(t_i,p_i,s_i),
\end{equation}
where $t_i$, $p_i$, and $s_i$ denote its type, scope, and subject.

%Generating multiple candidates allows the model to explore alternative interpretations of the same code change. It also reveals patterns repeatedly supported across generations: reliable types, scopes, and descriptions tend to recur, whereas uncertain predictions are more dispersed.

\subsection{Two-Dimensional Consistency Reranking}
\label{sec:two_dimensional_consistency}

The generated candidates are reranked using horizontal and vertical consistency. Horizontal consistency measures whether a subject agrees with the input commit and the structured fields within a message. Vertical consistency captures the support that a prediction receives from other candidates generated for the same commit.

\subsubsection{Vertical Structural Consensus}
\label{sec:structural_consensus}

We first derive consensus predictions for the two structural fields. The most frequent type and scope among the $N$ candidates are computed as
\begin{equation}
\hat{r}
=
\arg\max_{v}
\sum_{i=1}^{N}
\mathbb{I}(r_i=v),
\qquad r\in\{t,p\},
\end{equation}
where $\mathbb{I}(\cdot)$ is the indicator function, $r_i$ is the type or scope of the $i$-th candidate, and $v$ is a possible field value.

The resulting $\hat{t}$ and $\hat{p}$ represent the structural consensus of the candidate set. Instead of retaining the original and potentially noisy structural fields of each candidate, we evaluate all candidate subjects under the shared consensus type and scope. This provides a unified structural context for subsequent consistency scoring.

\subsubsection{Horizontal Component Consistency}
\label{sec:horizontal_consistency}

For each candidate subject $s_i$, we construct a metadata sequence from the consensus type and scope:
\begin{equation}
z
=
\hat{t};
\hat{p}.
\end{equation}
A learned consistency scorer then assigns
\begin{equation}
q_i
=
f_{\theta}(x,z,s_i),
\end{equation}
where $q_i$ measures how well the subject describes the code change while remaining compatible with the predicted change category and affected scope.

%A high score therefore indicates that the commit, type, scope, and subject form a coherent description of the same modification. Incorporating the structural fields also allows the scorer to distinguish a generally relevant subject from one that is inconsistent with the predicted type or scope.

\subsubsection{Vertical Semantic Consensus}
\label{sec:semantic_consensus}

The candidate set often contains subjects that express the same
interpretation with minor lexical variations. Exact matching would treat
such paraphrases as distinct and thus underestimate their shared support,
a known limitation when aggregating free-form generations
\cite{chen2024universal,kuhn2023semantic, oh2026latent}. We therefore group weakly
matched subjects into $K$ clusters, denoted as
$C_1,C_2,\ldots,C_K$.

Before matching, we normalize case and whitespace, remove leading and trailing punctuation, and tokenize subjects using the same tokenizer as in dataset processing. Two subjects are considered a weak match if their token-length difference is at most three and their similarity score is at least $0.97$. Subjects satisfying both conditions are assigned to the same cluster.

For each cluster $C_k$, we aggregate the consistency scores of its members:
\begin{equation}
Q(C_k)
=
\sum_{i:s_i\in C_k} q_i.
\end{equation}
The aggregated score reflects both how frequently an interpretation appears and how well its realizations align with the commit and consensus structural fields. Thus, a cluster scores highly when its interpretation is both frequent and semantically reliable.

We rank clusters by their aggregated scores and select, from each ranked cluster $C_{(k)}$, the subject with the highest individual consistency score:
\begin{equation}
s_k^{*}
=
\arg\max_{s_i\in C_{(k)}} q_i.
\end{equation}
The selected subjects form the reranked list
$\mathcal{S}^{*}=[s_1^{*},s_2^{*},\ldots,s_K^{*}]$,
and the top-ranked subject is combined with the consensus type and scope to produce the final output:
\begin{equation}
m^{*}
=
\hat{t}(\hat{p}):s_1^{*}.
\end{equation}

In this procedure, the individual score $q_i$ captures horizontal consistency within a conventional message, while cluster-level aggregation captures vertical consistency across generations. %Selecting one representative from each cluster also reduces redundant outputs and preserves diversity in the reranked list.

\begin{table*}[t]
    \centering
    \small
    \begin{tabular}{llllccccc}
        \toprule
        \multirow{2}{*}{\textbf{Method}}
        & 
        & \multirow{2}{*}{\textbf{Category}}
        & \multirow{2}{*}{\textbf{Params}}
        & \multicolumn{1}{c}{\textbf{Type}}
        & \multicolumn{1}{c}{\textbf{Scope}}
        & \multicolumn{3}{c}{\textbf{Subject}} \\
        %\cmidrule(lr){4-4}
        \cmidrule(lr){5-5}
        \cmidrule(lr){6-6}
        \cmidrule(lr){7-9}
        &
        & 
        &
        & \textbf{Acc.}
        & \textbf{BLEU}
        & \textbf{BLEU}
        & \textbf{ROUGE-L}
        & \textbf{METEOR} \\
        \midrule

        NNGen~\cite{liu2018neural}
        & &  Retrieval-based  & -- & -- & -- & 22.74 & 13.32 & 6.98 \\
        CoDiSum~\cite{xu2019commit}
        & & Learning-based & -- & -- & -- & 18.50 & 10.01 & 4.54 \\
        CoreGen~\cite{nie2021coregen}
        & & Learning-based & -- & -- & -- & 15.88 & 8.85 & 5.04 \\
        FIRA~\cite{dong2022fira}
        & & Learning-based & -- & -- & -- & 23.97 & 20.03 & \textbf{12.48} \\
        RACE~\cite{shi2022race}
        & & Hybrid & -- & -- & -- & 22.80 & 18.72 & 11.87 \\
        COME~\cite{he2023come}
        & & Hybrid & -- & -- & -- & 22.60 & 13.74 & 9.21 \\

        \midrule

        \multirow{2}{*}{Qwen2.5-Coder}
        & Base & LLM-Decoder & 7B & 35.01 & 40.49 & 23.37 & 15.64 & 8.58 \\
        & \textsc{CoCoRerank} & LLM-Decoder & 7B & 37.43 & 42.94 & 24.48 & 17.64 & 9.78 \\

        \midrule

        \multirow{2}{*}{DeepSeek-Coder-V2}
        & Base & LLM-Decoder & 16B & 31.40 & 28.63 & 19.71 & 14.54 & 8.52 \\
        & \textsc{CoCoRerank} & LLM-Decoder & 16B & 39.57 & 29.14 & 21.03 & 16.84 & 10.05 \\

        \midrule

        \multirow{2}{*}{Meta-Llama-3.1}
        & Base & LLM-Decoder & 8B & 41.44 & 41.17 & 24.58 & 18.72 & 9.97 \\
        & \textsc{CoCoRerank} & LLM-Decoder & 8B & \textbf{48.52} & \textbf{43.18} & \textbf{25.91} & \textbf{20.06} & 11.25 \\

        \bottomrule
    \end{tabular}
    \begin{tablenotes}
        \footnotesize
        \item $^{*}$ "--" indicates that the corresponding method does not evaluate this component or the parameter scale is not officially reported.
    \end{tablenotes}

    \caption{Comparison with representative commit message generation baselines and large language models. "Base" denotes the original candidate order without reranking, with its Top-1 output corresponding to the first candidate in the list of 100 candidates.}
    \label{tab:main_results}
\end{table*}

\begin{table}[t]
    \centering
    \small
    \setlength{\tabcolsep}{0.8mm}
    \begin{tabular}{llccc}
        \toprule
        \textbf{Generator}
        & \textbf{Setting}
        & \textbf{Pass@5}
        & \textbf{Pass@2}
        & \textbf{Pass@1} \\
        \midrule

        \multirow{2}{*}{Qwen2.5-Coder} & Base
        & 8.66  & 7.77 & 6.96 \\
        & \textsc{CoCoRerank}
        & \textbf{9.83}  & \textbf{8.80} & \textbf{8.15} \\

        \midrule
        \multirow{2}{*}{DeepSeek-Coder-V2} & Base
        & 7.28  & 6.28 & 5.71 \\
        & \textsc{CoCoRerank}
        & \textbf{8.13}  & \textbf{7.73} & \textbf{7.35} \\

        \midrule
        \multirow{2}{*}{Meta-Llama-3.1} & Base
        & 11.26  & 9.64 & 8.49 \\
        & \textsc{CoCoRerank}
        & \textbf{12.96}  & \textbf{11.86} & \textbf{10.86} \\
        \bottomrule
    \end{tabular}
    \caption{Pass@$k$ performance of \textsc{CoCoRerank} across different
    large language model generators.}
    \label{tab:llm_reranking}
\end{table}

\subsection{Consistency Scorer}
\label{sec:consistency_scorer}

The consistency scorer, implemented with Qwen3-0.6B, evaluates the semantic compatibility among the code diff, consensus metadata, and candidate subject. Given a code diff $x$, metadata sequence $z$, and subject $s$, we construct the conventional context and subject representations as

\begin{equation}
\mathbf{h}_{c}
=
[E_x(x);E_z(z)],
\qquad
\mathbf{h}_{s}
=
E_s(s),
\end{equation}
where $E_x$, $E_z$, and $E_s$ denote the corresponding encoders.

The consistency score is then computed as
\begin{equation}
f_{\theta}(x,z,s)
=
\operatorname{sim}
\left(
g_c(\mathbf{h}_{c}),
g_s(\mathbf{h}_{s})
\right),
\end{equation}
where $g_c$ and $g_s$ project the representations into the same embedding space and $\operatorname{sim}(\cdot)$ denotes the cosine similarity.

%Unlike conventional diff--subject matching, the structured context explicitly incorporates type and scope. The resulting score therefore reflects not only whether the subject is relevant to the code change, but also whether it is compatible with the predicted change category and affected component.

We then train the consistency scorer in two stages. The first stage learns general alignment between conventional contexts and ground-truth subjects, while the second stage improves fine-grained discrimination among plausible subjects generated for the same commit.

\paragraph{Easy Stage}
Given a mini-batch of $B$ training instances
${(c_i,s_i^{+})}_{i=1}^{B}$, where $c_i=(x_i,z_i)$ and $s_i^{+}$ is the ground-truth subject, we adopt an in-batch multi-positive contrastive objective \cite{khosla2020supervised}. For brevity, we write $f_{\theta}(c_i,s)=f_{\theta}(x_i,z_i,s)$. Subjects with the same normalized ground-truth text are treated as positives, with
$P(i)=\{j\mid s_j^{+}=s_i^{+}\}$.

The context-to-subject loss is defined as
\begin{equation}
\mathcal{L}_{c\rightarrow s}
=
-\frac{1}{B}
\sum_{i=1}^{B}
\log
\frac{
\sum_{j\in P(i)}
\exp\!\left(f_{\theta}(c_i,s_j^{+})/\tau\!\right)
}{
\sum_{j=1}^{B}
\exp\!\left(f_{\theta}(c_i,s_j^{+})/\tau\!\right)
},
\end{equation}
where $\tau$ is the temperature parameter. The symmetric subject-to-context loss, $\mathcal{L}_{s\rightarrow c}$, is defined analogously. The easy-stage objective is
\begin{equation}
\mathcal{L}_{\mathrm{e}}
=
\frac{1}{2}
\left(
\mathcal{L}_{c\rightarrow s}
+
\mathcal{L}_{s\rightarrow c}
\right).
\end{equation}
The resulting model is then used to initialize the hard-negative training stage.

\paragraph{Hard Stage}
Random in-batch negatives are often easy to distinguish because they describe unrelated code changes. In the actual reranking scenario, however, the scorer must discriminate among multiple fluent and plausible subjects generated for the same commit. Motivated by the effectiveness of hard negatives in contrastive learning \cite{kalantidis2020hard,liu2025uncertainty}, we further introduce a hard-negative
training stage.

For each training instance, we construct a hard-negative set
$\mathcal{H}_i=
\{s_{i,1}^-,\ldots,s_{i,K_i}^-\}$.
These negatives are generated from the same code change and are therefore semantically close to the ground-truth subject, but may contain inaccurate, incomplete, or misleading descriptions. Candidates identical to the ground truth are removed to avoid false negatives.

The hard-negative loss is defined as
\begin{equation}
\begin{aligned}
\mathcal{L}_{\mathrm{h}}
&=
\frac{1}{B}
\sum_{i=1}^{B}
\log
\!\left[
1
+
\sum_{k=1}^{K_i}
\frac{
\exp\!\left(f_{\theta}(c_i,s_{i,k}^{-})/\tau_h\!\right)
}{
\exp\!\left(f_{\theta}(c_i,s_i^{+})/\tau_h\!\right)
}
\!\right],
\end{aligned}
\end{equation}
where $\tau_h$ is the temperature parameter for within-instance discrimination.

During this stage, we retain the in-batch multi-positive objective and jointly optimize
\begin{equation}
\mathcal{L}
=
\mathcal{L}_{\mathrm{e}}
+
\lambda\mathcal{L}_{\mathrm{h}},
\end{equation}
where $\lambda$ controls the contribution of hard negatives.

% \begin{table*}[t]
%     \centering
%     \small
%     \setlength{\tabcolsep}{3mm}
%     \begin{tabular}{lccc|ccc}
%         \toprule
%         \textbf{Method}
%         & \textbf{BLEU}
%         & \textbf{ROUGE-L}
%         & \textbf{METEOR}
%         & \textbf{Pass@5}
%         & \textbf{Pass@2}
%         & \textbf{Pass@1} \\
%         \midrule
%         \textbf{Full Method}
%         & \textbf{24.48} & \textbf{17.64} & \textbf{9.78}
%         & \textbf{9.83} 
%         & \textbf{8.80} & \textbf{8.15} \\
%         w/o Horizontal Consistency
%         & 24.08 & 16.18 & 8.88
%         & 9.44  & 8.15 & 7.35 \\
%         w/o Vertical Consistency
%         & 24.14 & 17.39 & 9.67
%         & 8.62  & 8.13 & 7.90 \\
%         w/o Hard Negatives
%         & 24.19 & 17.29 & 9.62
%         & 9.43  & 8.59 & 7.84 \\
%         Base
%         & 23.37 & 15.64 & 8.58
%         & 8.66  & 7.77 & 6.96 \\
%         \bottomrule
%     \end{tabular}
%     \caption{Ablation study of the proposed consistency reranking
%     framework.}
%     \label{tab:method_ablation}
% \end{table*}

\section{Experimental Setup}
\label{sec:experiments}

In this section, we conduct extensive experiments to evaluate the proposed dataset and \textsc{CoCoRerank}, including comparisons with representative CMG baselines, comparisons of different reranking strategies, and ablation studies on conventional generation and individual method components. Implementation details are provided in the Appendix.

% Specifically, we aim to
% answer the following research questions:

% \begin{itemize}
%     \item \textbf{RQ1:} How does \textsc{CoCoRerank} compare with representative
%     commit message generation baselines and large language models?
%     \item \textbf{RQ2:} How effective is \textsc{CoCoRerank}
%     compared with selected candidate reranking strategies?
%     \item \textbf{RQ3:} Can \textsc{CoCoRerank} generalize
%     across different code-oriented large language models?
%     \item \textbf{RQ4:} Does explicitly generating CCS-compliant
%     messages improve commit message generation?
%     \item \textbf{RQ5:} How much does each component of  \textsc{CoCoRerank} contribute?
% \end{itemize}

% \subsection{Experimental Setup}
\label{sec:experimental_setup}
 
\subsubsection{Baselines} We adopt three categories of advanced baselines, three instruction-tuned code large language models, as well as two typical reranking strategies for evaluation. 
\begin{itemize}
    \item \textbf{CMG Baselines:} representative retrieval-based, learning-based, and hybrid CMG approaches. 

    \item \textbf{LLM Generators:} Qwen2.5-Coder-7B-Instruct \cite{hui2024qwen25codertechnicalreport}, DeepSeek-Coder-V2-Lite-Instruct \cite{deepseekai2024deepseekcoderv2breakingbarrierclosedsource}, and Meta-Llama-3.1-8B-Instruct \cite{grattafiori2024llama3herdmodels}.

    \item \textbf{Reranking Strategies:} MBR \cite{kumar2004minimum} computes pairwise candidate similarities and selects candidates with the highest aggregate similarity to the rest of the set. Majority Voting ranks exact duplicate candidates according to their occurrence frequencies.
\end{itemize}

%\begin{itemize}
%    \item \textbf{CMG Baselines:} We compare our method with representative retrieval-based, learning-based, and Hybrid CMG approaches.
%    \item \textbf{LLM Generators:} We evaluate three instruction-tuned large language models: Qwen2.5-Coder-7B-Instruct \cite{hui2024qwen25codertechnicalreport}, DeepSeek-Coder-V2-Lite-Instruct \cite{deepseekai2024deepseekcoderv2breakingbarrierclosedsource}, and Meta-Llama-3.1-8B-Instruct \cite{grattafiori2024llama3herdmodels}. 
%    \item \textbf{Reranking Strategies:} We compare our method with two representative reranking strategies. MBR \cite{kumar2004minimum} computes pairwise candidate similarities and selects candidates with the highest aggregate similarity to the rest of the set. Majority Voting ranks exact duplicate candidates according to their occurrence frequencies.
%\end{itemize}

\subsubsection{Evaluation Metrics}

We evaluate three distinct components of a conventional message separately. Type prediction is measured by accuracy.
Scope generation is evaluated using BLEU \cite{bleu}, with common file-name and path separators (e.g., "-" and "/") treated as token delimiters.
Subject generation is
evaluated using BLEU, ROUGE-L \cite{rouge}, and METEOR \cite{meteor}. %Higher values indicate better agreement with the human-written references.

We further report $\mathrm{Pass@}k$ to measure whether the top-$k$
reranked candidates contain a high-quality subject. For the $i$-th test
instance, a hit is defined as
\begin{equation}
    \mathrm{hit}_i(s)
    =
    \mathbb{I}
    \left(
        \mathrm{BLEU}(s,s_i^{*}) \geq 70
    \right).
\end{equation}
The $\mathrm{Pass@}k$ score is then defined as
\begin{equation}
    \mathrm{Pass@}k
    =
    \frac{1}{|\mathcal{D}|}
    \sum_{i=1}^{|\mathcal{D}|}
    \max_{j\leq k}
    \mathrm{hit}_i(\hat{s}_{i,j}).
\end{equation}
Unlike corpus-level overlap metrics, $\mathrm{Pass@}k$ directly measures
whether reranking places a highly matching candidate near the top of the
list.

\begin{table}[t]
    \centering
    \small
    \setlength{\tabcolsep}{2.5mm}
    \begin{tabular}{lccc}
        \toprule
        %\multicolumn{4}{c}{\textbf{Generation Quality}} \\
        %\midrule
        \textbf{Method}
        & \textbf{BLEU}
        & \textbf{ROUGE-L}
        & \textbf{METEOR} \\
        \midrule
        Base
        & 23.37 & 15.64 & 8.58 \\
        MBR
        & 23.70 & 16.32 & 9.01 \\
        Majority Voting
        & 24.11 & 16.17 & 8.85 \\
        \textbf{\textsc{CoCoRerank}}
        & \textbf{24.48}
        & \textbf{17.64}
        & \textbf{9.78} \\

        \midrule
        %\multicolumn{4}{c}{\textbf{Ranking Performance}} \\
        %\midrule
        \textbf{Method}
        & \textbf{Pass@5}
        & \textbf{Pass@2}
        & \textbf{Pass@1} \\
        \midrule
        Base
        & 8.66 & 7.77 & 6.96 \\
        MBR
        & 9.23 & 8.10 & 7.30 \\
        Majority Voting
        & 7.38 & 7.37 & 7.37 \\
        \textbf{\textsc{CoCoRerank}}
        & \textbf{9.83}
        & \textbf{8.80}
        & \textbf{8.15} \\
        \bottomrule
    \end{tabular}
    \caption{Comparison with selected reranking strategies.}
    \label{tab:reranking_results}
\end{table}

\begin{table*}[t]
    \centering
    \small
    \setlength{\tabcolsep}{2.0mm}
    \begin{tabular}{l *{6}{c}}
        \toprule
        \textbf{Method}
        & \textbf{BLEU}
        & \textbf{ROUGE-L}
        & \textbf{METEOR}
        & \textbf{Pass@5}
        & \textbf{Pass@2}
        & \textbf{Pass@1} \\
        \midrule
        \textbf{Full Method}
        & 24.48 & 17.64 & 9.78
        & 9.83  & 8.80  & 8.15 \\
        w/o Horizontal Consistency
        & \textbf{24.08} \tiny{(-1.63\%)} & \textbf{16.18} \tiny{(-8.28\%)} & \textbf{8.88} \tiny{(-9.20\%)}
        & 9.44 \tiny{(-3.97\%)}  & 8.15 \tiny{(-7.39\%)}  & \textbf{7.35} \tiny{(-9.82\%)} \\
        w/o Vertical Consistency
        & 24.14 \tiny{(-1.39\%)} & 17.39 \tiny{(-1.42\%)} & 9.67 \tiny{(-1.12\%)}
        & \textbf{8.62} \tiny{(-12.31\%)} & \textbf{8.13} \tiny{(-7.61\%)}  & 7.90 \tiny{(-3.07\%)} \\
        w/o Hard Negatives
        & 24.19 \tiny{(-1.18\%)} & 17.29 \tiny{(-1.98\%)} & 9.62 \tiny{(-1.64\%)}
        & 9.43 \tiny{(-4.07\%)}  & 8.59 \tiny{(-2.39\%)}  & 7.84 \tiny{(-3.80\%)} \\
        % Base
        % & 23.37 \tiny{(-4.53\%)} & 15.64 \tiny{(-11.34\%)} & 8.58 \tiny{(-12.27\%)}
        % & 8.66 \tiny{(-11.90\%)} & 7.77 \tiny{(-11.70\%)}  & 6.96 \tiny{(-14.60\%)} \\
        \bottomrule
    \end{tabular}
    \begin{tablenotes}
        \footnotesize
        \item $^{*}$ The tiny values in parentheses denote the relative performance drop ratio (\%) compared with the full method.
    \end{tablenotes}
    \caption{Ablation study of the proposed consistency reranking framework.}
    \label{tab:method_ablation}
\end{table*}

\begin{table}[t]
    \centering
    \small
    \setlength{\tabcolsep}{1mm}
    \begin{tabular}{llccc}
        \toprule
        \textbf{Generator}
        & \textbf{Setting}
        & \textbf{BLEU}
        & \textbf{ROUGE-L}
        & \textbf{METEOR} \\
        \midrule

        \multirow{2}{*}{Qwen2.5-Coder}
        & Subject
        & 20.86 & 14.76 & 8.25 \\
        & CCS
        & \textbf{23.37} & \textbf{15.64} & \textbf{8.58} \\

        \midrule
        \multirow{2}{*}{DeepSeek-Coder-V2}
        & Subject
        & 17.09 & 12.42 & 7.98 \\
        & CCS
        & \textbf{19.71} & \textbf{14.54} & \textbf{8.52} \\

        \midrule
        \multirow{2}{*}{Meta-Llama-3.1}
        & Subject
        & 24.21 & 17.46 & 9.36 \\
        & CCS
        & \textbf{24.58} & \textbf{18.72} & \textbf{9.97} \\

        \bottomrule
    \end{tabular}
    \caption{Ablation study of CCS-structured generation.}
    \label{tab:ccs_ablation}
\end{table}

\section{Experimental Results}

\subsection{Comparison of CMG Baselines}
\label{sec:overall_comparison}

We compare \textsc{CoCoRerank} with representative CMG baselines and evaluate it across three different code-oriented LLM generators. Table~\ref{tab:main_results} reports the detailed component-level generation results, while Table~\ref{tab:llm_reranking} presents the corresponding overall $\mathrm{Pass@}k$ performance.

As shown in Table~\ref{tab:main_results}, applying \textsc{CoCoRerank} consistently improves type, scope, and subject generation across all three underlying LLMs. Among the subject metrics, METEOR shows the largest relative gains, increasing by 13.99\%, 17.96\%, and 12.84\% for Qwen2.5-Coder, DeepSeek-Coder-V2, and Meta-Llama-3.1, respectively. The strongest overall performance is achieved with Meta-Llama-3.1-8B-Instruct, reaching 48.52\% type accuracy, 43.18 scope BLEU, 25.91 subject BLEU, and 20.06 subject ROUGE-L. Nevertheless, FIRA retains the highest METEOR score, indicating that the performance ceiling of reranking remains constrained by the candidate-generation capability of the underlying model.

The improvements are also reflected in candidate ranking. As shown in Table~\ref{tab:llm_reranking}, \textsc{CoCoRerank} consistently improves all $\mathrm{Pass@}k$ metrics across the three generators. In particular, $\mathrm{Pass@1}$ improves by 17.10\%, 28.72\%, and 27.92\% for Qwen2.5-Coder, DeepSeek-Coder-V2, and Meta-Llama-3.1, respectively.
These consistent gains in both component-level generation and candidate ranking demonstrate the effectiveness of \textsc{CoCoRerank} and its generalizability across different LLM generators.

\subsection{Comparison of Reranking Strategies}
\label{sec:reranking_comparison}

We compare \textsc{CoCoRerank} with MBR and Majority Voting on candidates generated by Qwen2.5-Coder-7B-Instruct. As shown in Table~\ref{tab:reranking_results}, MBR improves over the base output across all metrics, confirming that agreement among multiple candidates provides useful ranking signals. Majority Voting also improves the subject-generation metrics and $\mathrm{Pass@1}$, but yields lower $\mathrm{Pass@5}$ and $\mathrm{Pass@2}$, reflecting the limitations of relying solely on exact candidate frequency. In comparison, MBR captures inter-candidate similarity but does not explicitly measure consistency with the code change and structured fields.

By jointly exploiting candidate consensus and commit-conditioned consistency, \textsc{CoCoRerank} achieves the best performance across all subject-generation and $\mathrm{Pass@}k$ metrics. Its weak-match clustering captures semantic agreement beyond exact duplicates, while the consistency scorer filters out frequent but commit-inconsistent candidates. Relative to the base output, \textsc{CoCoRerank} yields gains of 4.75\% in BLEU, 12.79\% in ROUGE-L, 13.99\% in METEOR, and 17.10\% in $\mathrm{Pass@1}$. These results demonstrate that combining the two consistency signals provides more effective candidate selection than either similarity-based or frequency-based reranking alone.

\subsection{Ablation Study}
\label{sec:ablation}

We conduct ablation experiments from two perspectives: the effect of conventional generation and the contribution of individual components in \textsc{CoCoRerank}.

We first examine whether explicitly generating the complete CCS structure benefits subject generation. We compare subject-only prompting with conventional prompting across three LLM generators. As shown in Table~\ref{tab:ccs_ablation}, conventional generation consistently improves BLEU, ROUGE-L, and METEOR across all three generators. Averaged across the three generators, ROUGE-L achieves the largest relative improvement of 10.08\%. Among all three  generators, \mbox{DeepSeek-Coder-V2} benefits the most, with an average relative gain of 13.06\% across the three metrics. This suggests that jointly generating type and scope provides useful structural guidance for producing higher-quality subjects.

We then analyze the contribution of each component in \textsc{CoCoRerank} using Qwen2.5-Coder-7B-Instruct as the underlying generator. As shown in Table~\ref{tab:method_ablation}, removing any component degrades performance, while different components exhibit distinct effects across metrics. Removing horizontal consistency causes the largest degradation in BLEU (1.63\%), ROUGE-L (8.28\%), METEOR (9.20\%), and $\mathrm{Pass@1}$ (9.82\%), highlighting its importance for selecting a high-quality final output. In contrast, removing vertical consistency has a greater impact on $\mathrm{Pass@5}$ (12.31\%) and $\mathrm{Pass@2}$ (7.61\%), indicating that candidate-level consensus is particularly important for preserving high-quality candidates among the top-ranked outputs. Removing hard-negative training also consistently degrades all metrics, demonstrating its contribution to the overall discriminative ability of the consistency scorer.

Together, these results demonstrate that explicitly generating the complete CCS structure provides a strong foundation for conventional CMG, while all components play complementary roles in \textsc{CoCoRerank}.

\section{Conclusion}
\label{sec:conclusion}

This paper studies conventional commit message generation under the Conventional Commits Specification. We construct a high-quality dataset of 86,688 commits that preserves the complete \typeword{type} (\scopeword{scope}): \subjectword{subject} structure and enables component-level evaluation. We further propose a two-dimensional consistency-based reranking framework for LLM-based conventional commit message generation. Experiments show that the framework improves both structural component prediction and subject generation quality.

In the future, we plan to further investigate how the structural components of commit messages, including \typeword{type}, \scopeword{scope}, and \subjectword{subject}, are reflected in code changes, with the goal of developing more structure-aware generation methods that better capture the semantic roles of message components.

\clearpage

\bibliography{aaai2027}

@inproceedings{buse2010automatically,
  title={Automatically documenting program changes},
  author={Buse, Raymond PL and Weimer, Westley R},
  booktitle={Proceedings of the 25th IEEE/ACM international conference on automated software engineering},
  pages={33--42},
  year={2010}
}

@inproceedings{cortescoy2014changescribe,
  title={On automatically generating commit messages via summarization of source code changes},
  author={Cort{\'e}s-Coy, Luis Fernando and Linares-V{\'a}squez, Mario and Aponte, Jairo and Poshyvanyk, Denys},
  booktitle={2014 IEEE 14th International Working Conference on Source Code Analysis and Manipulation},
  pages={275--284},
  year={2014},
  organization={IEEE}
}

@inproceedings{shen2016automatic,
  title={On automatic summarization of what and why information in source code changes},
  author={Shen, Jinfeng and Sun, Xiaobing and Li, Bin and Yang, Hui and Hu, Jiajun},
  booktitle={2016 IEEE 40th Annual Computer Software and Applications Conference (COMPSAC)},
  volume={1},
  pages={103--112},
  year={2016},
  organization={IEEE}
}

@inproceedings{liu2018neural,
  title={Neural-machine-translation-based commit message generation: how far are we?},
  author={Liu, Zhongxin and Xia, Xin and Hassan, Ahmed E and Lo, David and Xing, Zhenchang and Wang, Xinyu},
  booktitle={Proceedings of the 33rd ACM/IEEE international conference on automated software engineering},
  pages={373--384},
  year={2018}
}

@inproceedings{hoang2020cc2vec,
  title={Cc2vec: Distributed representations of code changes},
  author={Hoang, Thong and Kang, Hong Jin and Lo, David and Lawall, Julia},
  booktitle={Proceedings of the ACM/IEEE 42nd international conference on software engineering},
  pages={518--529},
  year={2020}
}

@inproceedings{jiang2017automatically,
  title={Automatically generating commit messages from diffs using neural machine translation},
  author={Jiang, Siyuan and Armaly, Ameer and McMillan, Collin},
  booktitle={2017 32nd IEEE/ACM International Conference on Automated Software Engineering (ASE)},
  pages={135--146},
  year={2017},
  organization={IEEE}
}

@inproceedings{xu2019commit,
  author       = {Shengbin Xu and
                  Yuan Yao and
                  Feng Xu and
                  Tianxiao Gu and
                  Hanghang Tong and
                  Jian Lu},
  editor       = {Sarit Kraus},
  title        = {Commit Message Generation for Source Code Changes},
  booktitle    = {Proceedings of the Twenty-Eighth International Joint Conference on
                  Artificial Intelligence, {IJCAI} 2019, Macao, China, August 10-16,
                  2019},
  pages        = {3975--3981},
  publisher    = {ijcai.org},
  year         = {2019},
  url          = {https://doi.org/10.24963/ijcai.2019/552},
  doi          = {10.24963/IJCAI.2019/552},
  bibsource    = {dblp computer science bibliography, https://dblp.org}
}

@article{nie2021coregen,
  title={Coregen: Contextualized code representation learning for commit message generation},
  author={Nie, Lun Yiu and Gao, Cuiyun and Zhong, Zhicong and Lam, Wai and Liu, Yang and Xu, Zenglin},
  journal={Neurocomputing},
  volume={459},
  pages={97--107},
  year={2021},
  publisher={Elsevier}
}

@inproceedings{dong2022fira,
  title={Fira: fine-grained graph-based code change representation for automated commit message generation},
  author={Dong, Jinhao and Lou, Yiling and Zhu, Qihao and Sun, Zeyu and Li, Zhilin and Zhang, Wenjie and Hao, Dan},
  booktitle={Proceedings of the 44th international conference on software engineering},
  pages={970--981},
  year={2022}
}

@inproceedings{he2023come,
  title={Come: Commit message generation with modification embedding},
  author={He, Yichen and Wang, Liran and Wang, Kaiyi and Zhang, Yupeng and Zhang, Hang and Li, Zhoujun},
  booktitle={Proceedings of the 32nd ACM SIGSOFT International Symposium on Software Testing and Analysis},
  pages={792--803},
  year={2023}
}

@article{liu2020atom,
  title={Atom: Commit message generation based on abstract syntax tree and hybrid ranking},
  author={Liu, Shangqing and Gao, Cuiyun and Chen, Sen and Nie, Lun Yiu and Liu, Yang},
  journal={IEEE Transactions on Software Engineering},
  volume={48},
  number={5},
  pages={1800--1817},
  year={2020},
  publisher={IEEE}
}

@inproceedings{shi2022race,
  title={Race: Retrieval-augmented commit message generation},
  author={Shi, Ensheng and Wang, Yanlin and Tao, Wei and Du, Lun and Zhang, Hongyu and Han, Shi and Zhang, Dongmei and Sun, Hongbin},
  booktitle={Proceedings of the 2022 Conference on Empirical Methods in Natural Language Processing},
  pages={5520--5530},
  year={2022}
}

@inproceedings{tao2021evaluation,
  title={On the evaluation of commit message generation models: An experimental study},
  author={Tao, Wei and Wang, Yanlin and Shi, Ensheng and Du, Lun and Han, Shi and Zhang, Hongyu and Zhang, Dongmei and Zhang, Wenqiang},
  booktitle={2021 IEEE International Conference on Software Maintenance and Evolution (ICSME)},
  pages={126--136},
  year={2021},
  organization={IEEE}
}

@inproceedings{schall2024commitbench,
  title={Commitbench: A benchmark for commit message generation},
  author={Schall, Maximilian and Czinczoll, Tamara and De Melo, Gerard},
  booktitle={2024 IEEE International Conference on Software Analysis, Evolution and Reengineering (SANER)},
  pages={728--739},
  year={2024},
  organization={IEEE}
}

@inproceedings{liu2019generating,
  title={Generating commit messages from diffs using pointer-generator network},
  author={Liu, Qin and Liu, Zihe and Zhu, Hongming and Fan, Hongfei and Du, Bowen and Qian, Yu},
  booktitle={2019 IEEE/ACM 16th International Conference on Mining Software Repositories (MSR)},
  pages={299--309},
  year={2019},
  organization={IEEE}
}

@INPROCEEDINGS{Bar2016relation,
  author={Barnett, Jacob G. and Gathuru, Charles K. and Soldano, Luke S. and McIntosh, Shane},
  booktitle={2016 IEEE/ACM 13th Working Conference on Mining Software Repositories (MSR)}, 
  title={The Relationship between Commit Message Detail and Defect Proneness in Java Projects on GitHub}, 
  year={2016},
  volume={},
  number={},
  pages={496-499},
  doi={}}

@INPROCEEDINGS{Li2023matter,
  author={Li, Jiawei and Ahmed, Iftekhar},
  booktitle={2023 IEEE/ACM 45th International Conference on Software Engineering (ICSE)}, 
  title={Commit Message Matters: Investigating Impact and Evolution of Commit Message Quality}, 
  year={2023},
  volume={},
  number={},
  pages={806-817},
  doi={10.1109/ICSE48619.2023.00076}}

@inproceedings{bleu,
    title = "{B}leu: a Method for Automatic Evaluation of Machine Translation",
    author = "Papineni, Kishore  and
      Roukos, Salim  and
      Ward, Todd  and
      Zhu, Wei-Jing",
    editor = "Isabelle, Pierre  and
      Charniak, Eugene  and
      Lin, Dekang",
    booktitle = "Proceedings of the 40th Annual Meeting of the Association for Computational Linguistics",
    month = jul,
    year = "2002",
    address = "Philadelphia, Pennsylvania, USA",
    publisher = "Association for Computational Linguistics",
    url = "https://aclanthology.org/P02-1040/",
    doi = "10.3115/1073083.1073135",
    pages = "311--318"
}

@inproceedings{rouge,
    title = "{ROUGE}: A Package for Automatic Evaluation of Summaries",
    author = "Lin, Chin-Yew",
    booktitle = "Text Summarization Branches Out",
    month = jul,
    year = "2004",
    address = "Barcelona, Spain",
    publisher = "Association for Computational Linguistics",
    url = "https://aclanthology.org/W04-1013/",
    pages = "74--81"
}

@inproceedings{meteor,
    title = "{METEOR}: An Automatic Metric for {MT} Evaluation with Improved Correlation with Human Judgments",
    author = "Banerjee, Satanjeev  and
      Lavie, Alon",
    editor = "Goldstein, Jade  and
      Lavie, Alon  and
      Lin, Chin-Yew  and
      Voss, Clare",
    booktitle = "Proceedings of the {ACL} Workshop on Intrinsic and Extrinsic Evaluation Measures for Machine Translation and/or Summarization",
    month = jun,
    year = "2005",
    address = "Ann Arbor, Michigan",
    publisher = "Association for Computational Linguistics",
    url = "https://aclanthology.org/W05-0909/",
    pages = "65--72"
}

@INPROCEEDINGS{Zeng2025look,
  author={Zeng, Qunhong and Zhang, Yuxia and Qiu, Zhiqing and Liu, Hui},
  booktitle={2025 IEEE/ACM 47th International Conference on Software Engineering (ICSE)}, 
  title={A First Look at Conventional Commits Classification}, 
  year={2025},
  volume={},
  number={},
  pages={2277-2289},
  doi={10.1109/ICSE55347.2025.00011}}

@misc{wan2026commitsuitecomprehensivebenchmarkcommit,
      title={CommitSuite: A Comprehensive Benchmark for Commit Classification and Message Generation}, 
      author={Zirui Wan and Zhaonan Wu and Xinyi Hou and Yanjie Zhao and Pengcheng Xia and Haoyu Wang},
      year={2026},
      eprint={2605.02256},
      archivePrefix={arXiv},
}

@inproceedings{oh2026latent,
  title={Latent self-consistency for reliable majority-set selection in short-and long-answer reasoning},
  author={Oh, Jungsuk and Lee, Jay-Yoon},
  booktitle={Proceedings of the AAAI Conference on Artificial Intelligence},
  volume={40},
  number={38},
  pages={32591--32599},
  year={2026}
}

@inproceedings{liu2025uncertainty,
  title={Uncertainty-aware contrastive learning with hard negative sampling for code search tasks},
  author={Liu, Han and Zhan, Jiaqing and Zhang, Qin},
  booktitle={Proceedings of the AAAI Conference on Artificial Intelligence},
  volume={39},
  number={18},
  pages={18807--18815},
  year={2025}
}

@article{khosla2020supervised,
  title={Supervised contrastive learning},
  author={Khosla, Prannay and Teterwak, Piotr and Wang, Chen and Sarna, Aaron and Tian, Yonglong and Isola, Phillip and Maschinot, Aaron and Liu, Ce and Krishnan, Dilip},
  journal={Advances in neural information processing systems},
  volume={33},
  pages={18661--18673},
  year={2020}
}

@article{kalantidis2020hard,
  title={Hard negative mixing for contrastive learning},
  author={Kalantidis, Yannis and Sariyildiz, Mert Bulent and Pion, Noe and Weinzaepfel, Philippe and Larlus, Diane},
  journal={Advances in neural information processing systems},
  volume={33},
  pages={21798--21809},
  year={2020}
}

@article{li2022competition,
  title={Competition-level code generation with alphacode},
  author={Li, Yujia and Choi, David and Chung, Junyoung and Kushman, Nate and Schrittwieser, Julian and Leblond, R{\'e}mi and Eccles, Tom and Keeling, James and Gimeno, Felix and Dal Lago, Agustin and others},
  journal={Science},
  volume={378},
  number={6624},
  pages={1092--1097},
  year={2022},
  publisher={American Association for the Advancement of Science}
}

@inproceedings{
chen2023codet,
title={CodeT:  Code Generation with Generated Tests},
author={Bei Chen and Fengji Zhang and Anh Nguyen and Daoguang Zan and Zeqi Lin and Jian-Guang Lou and Weizhu Chen},
booktitle={The Eleventh International Conference on Learning Representations },
year={2023},
url={https://openreview.net/forum?id=ktrw68Cmu9c}
}

@inproceedings{kumar2004minimum,
  title={Minimum bayes-risk decoding for statistical machine translation},
  author={Kumar, Shankar and Byrne, Bill},
  booktitle={Proceedings of the Human Language Technology Conference of the North American Chapter of the Association for Computational Linguistics: HLT-NAACL 2004},
  pages={169--176},
  year={2004}
}

@inproceedings{
chen2024universal,
title={Universal Self-Consistency for Large Language Models},
author={Xinyun Chen and Renat Aksitov and Uri Alon and Jie Ren and Kefan Xiao and Pengcheng Yin and Sushant Prakash and Charles Sutton and Xuezhi Wang and Denny Zhou},
booktitle={ICML 2024 Workshop on In-Context Learning},
year={2024},
url={https://openreview.net/forum?id=LjsjHF7nAN}
}

@inproceedings{
kuhn2023semantic,
title={Semantic Uncertainty: Linguistic Invariances for Uncertainty Estimation in Natural Language Generation},
author={Lorenz Kuhn and Yarin Gal and Sebastian Farquhar},
booktitle={The Eleventh International Conference on Learning Representations },
year={2023},
url={https://openreview.net/forum?id=VD-AYtP0dve}
}

@misc{hui2024qwen25codertechnicalreport,
      title={Qwen2.5-Coder Technical Report}, 
      author={Binyuan Hui and Jian Yang and Zeyu Cui and Jiaxi Yang and Dayiheng Liu and Lei Zhang and Tianyu Liu and Jiajun Zhang and Bowen Yu and Keming Lu and others},
      year={2024},
      eprint={2409.12186},
      archivePrefix={arXiv},
      primaryClass={cs.CL},
      url={https://arxiv.org/abs/2409.12186}, 
}

@misc{deepseekai2024deepseekcoderv2breakingbarrierclosedsource,
      title={DeepSeek-Coder-V2: Breaking the Barrier of Closed-Source Models in Code Intelligence}, 
      author={DeepSeek-AI and Qihao Zhu and Daya Guo and Zhihong Shao and Dejian Yang and Peiyi Wang and Runxin Xu and Y. Wu and Yukun Li and Huazuo Gao and others },
      year={2024},
      eprint={2406.11931},
      archivePrefix={arXiv},
      primaryClass={cs.SE},
      url={https://arxiv.org/abs/2406.11931}, 
}

@misc{grattafiori2024llama3herdmodels,
      title={The Llama 3 Herd of Models}, 
      author={Aaron Grattafiori and Abhimanyu Dubey and Abhinav Jauhri and Abhinav Pandey and Abhishek Kadian and Ahmad Al-Dahle and Aiesha Letman and Akhil Mathur and Alan Schelten and Alex Vaughan and others},
      year={2024},
      eprint={2407.21783},
      archivePrefix={arXiv},
      primaryClass={cs.AI},
      url={https://arxiv.org/abs/2407.21783}, 
}

@inproceedings{liu2026dynamic,
  title={Dynamic-static synergistic selection method for candidate code solutions with generated test cases},
  author={Liu, Ren-Biao and Xue, Jiang-Tian and Ma, Chao-Zeng and Sun, Hui and Li, Xin-Ye and Li, Ming},
  booktitle={Proceedings of the AAAI Conference on Artificial Intelligence},
  volume={40},
  number={38},
  pages={32096--32104},
  year={2026}
}

@article{sun2023enhancing,
  title={Enhancing unsupervised domain adaptation by exploiting the conceptual consistency of multiple self-supervised tasks},
  author={Sun, Hui and Li, Ming},
  journal={Science China Information Sciences},
  volume={66},
  number={4},
  pages={142101},
  year={2023},
  publisher={Springer}
}

@inproceedings{huang2024enhancing,
  title={Enhancing large language models in coding through multi-perspective self-consistency},
  author={Huang, Baizhou and Lu, Shuai and Wan, Xiaojun and Duan, Nan},
  booktitle={Proceedings of the 62nd Annual Meeting of the Association for Computational Linguistics (Volume 1: Long Papers)},
  pages={1429--1450},
  year={2024}
}

@inproceedings{
wang2023selfconsistency,
title={Self-Consistency Improves Chain of Thought Reasoning in Language Models},
author={Xuezhi Wang and Jason Wei and Dale Schuurmans and Quoc V Le and Ed H. Chi and Sharan Narang and Aakanksha Chowdhery and Denny Zhou},
booktitle={The Eleventh International Conference on Learning Representations },
year={2023},
url={https://openreview.net/forum?id=1PL1NIMMrw}
}

@article{du2025capturing,
  title={Capturing the context-aware code change via dynamic control flow graph for commit message generation},
  author={Du, Yali and Li, Ying and Ma, Yi-Fan and Li, Ming},
  journal={Machine Learning},
  volume={114},
  number={4},
  pages={94},
  year={2025},
}

@inproceedings{du2023pre,
  title={Pre-training code representation with semantic flow graph for effective bug localization},
  author={Du, Yali and Yu, Zhongxing},
  booktitle={Proceedings of the 31st ACM joint European software engineering conference and symposium on the foundations of software engineering},
  pages={579--591},
  year={2023}
}

@inproceedings{ma2023capturing,
  title={Capturing the Long-Distance Dependency in the Control Flow Graph via Structural-Guided Attention for Bug Localization.},
  author={Ma, Yi-Fan and Du, Yali and Li, Ming},
  booktitle={IJCAI},
  pages={2242--2250},
  year={2023}
}

@inproceedings{du2024joint,
  title={A joint learning model with variational interaction for multilingual program translation},
  author={Du, Yali and Sun, Hui and Li, Ming},
  booktitle={Proceedings of the 39th IEEE/ACM International Conference on Automated Software Engineering},
  pages={1907--1918},
  year={2024}
}

@article{du2025post,
  title={Post-incorporating code structural knowledge into pretrained models via icl for code translation},
  author={Du, Yali and Sun, Hui and Li, Ming},
  journal={IEEE Transactions on Software Engineering},
  year={2025},
}

@inproceedings{du2023beyond,
  title={Beyond lexical consistency: Preserving semantic consistency for program translation},
  author={Du, Yali and Ma, Yi-Fan and Xie, Zheng and Li, Ming},
  booktitle={2023 IEEE International Conference on Data Mining (ICDM)},
  pages={91--100},
  year={2023},
}

@inproceedings{liurandom,
  title={Random Selection Reveals Implicit Knowledge Consensus in Code Generation},
  author={Liu, Ren-Biao and Li, Xin-Ye and Sun, Hui and Du, Yali and Xue, Jiang-Tian and Li, Ming},
  booktitle={Forty-third International Conference on Machine Learning},
  year={2026}
}

@article{sun2026aces,
  title={ACES: Who Tests the Tests? Leave-One-Out AUC Consistency for Code Generation},
  author={Sun, Hui and Zhang, Yun-Ji and Xie, Zheng and Liu, Ren-Biao and Du, Yali and Li, Xin-Ye and Li, Ming},
  journal={arXiv preprint arXiv:2604.03922},
  year={2026}
}

@article{du2026design,
  title={Design-Specification Tiling for ICL-based CAD Code Generation},
  author={Du, Yali and Xi, San-Zhuo and Sun, Hui and Li, Ming},
  journal={arXiv preprint arXiv:2603.12712},
  year={2026}
}

@article{du2026cit,
  title={CIT-CAD: Constraint Intent Tree-based CAD Code Generation and Verification},
  author={Du, Yali and Sun, Hui and Xi, San-Zhuo and Li, Ming},
  journal={arXiv preprint arXiv:2609.07434},
  year={2026}
}

@article{crupi2025effectiveness,
  title={On the effectiveness of llm-as-a-judge for code generation and summarization},
  author={Crupi, Giuseppe and Tufano, Rosalia and Velasco, Alejandro and Mastropaolo, Antonio and Poshyvanyk, Denys and Bavota, Gabriele},
  journal={IEEE Transactions on Software Engineering},
  volume={51},
  number={8},
  pages={2329--2345},
  year={2025},
}

@article{virk2025calibration,
  title={Calibration of large language models on code summarization},
  author={Virk, Yuvraj and Devanbu, Premkumar and Ahmed, Toufique},
  journal={Proceedings of the ACM on Software Engineering},
  volume={2},
  number={FSE},
  pages={2944--2964},
  year={2025},
}


% \appendix
\twocolumn[
  \begin{@twocolumnfalse}
    \begin{center}
      \fontsize{18}{20}\selectfont\bfseries
          Appendix
      \vspace{1em}
    \end{center}
  \end{@twocolumnfalse}
]

% \documentclass[letterpaper]{article}

% %\usepackage[submission]{aaai2027}
% \usepackage{listings}
% \usepackage[
%     top=1in,
%     bottom=1in,
%     left=1in,
%     right=1in
% ]{geometry}
% \lstdefinestyle{prompt}{
%     basicstyle=\ttfamily\small,
%     numbers=left,
%     numberstyle=\scriptsize,
%     numbersep=6pt,
%     stepnumber=1,
%     breaklines=true,
%     breakatwhitespace=false,
%     columns=fullflexible,
%     keepspaces=true,
%     showstringspaces=false,
%     xleftmargin=1.5em
% }

% \renewcommand{\thesection}{\Alph{section}}

% \pagestyle{empty}

% \begin{document}
% \onecolumn
% \thispagestyle{empty}

\section{Dataset Construction Pipeline}
\label{appendix:dataset_details}
Here we provide the detailed step-by-step filtering procedures, precise data volume changes, and label normalization rules omitted from the main text.

\subsection{Detailed Data Collection and Preliminary Filtering}
We first select 206 GitHub repositories that consistently follow the Conventional Commits Specification. Repositories with fewer than 1,000 commits are discarded to avoid unrepresentative development histories. For each qualified repository, we extract code diffs, commit messages, and repository-level metadata.

We remove mechanically generated commits, including version updates, build synchronization, and automatic artifact maintenance commits, as such samples only contain trivial repetitive changes and contribute little to semantic learning. We further adopt a V-DO subject filter to retain action-oriented commit descriptions.

For token-based filtering, we strictly retain diffs ranging from 64 to 32,767 tokens and commit subjects ranging from 4 to 63 tokens. This strategy filters overly trivial diffs lacking semantics and ultra-long diffs that introduce computational redundancy, while eliminating overly short meaningless subjects and excessively verbose descriptions. After preliminary filtering, the dataset retains 96,453 instances.

\subsection{Detailed Structural Normalization Steps}
After preliminary screening, we conduct fine-grained structural cleaning. We first parse each commit message into type, scope, and subject fields and remove samples with missing type or scope fields, reducing the dataset scale from 96,453 to 96,130 instances.

We then standardize type vocabularies by eliminating uninterpretable arbitrary labels and merging misspelled, abbreviated, and low-frequency labels into unified standard categories. This process effectively alleviates label fragmentation and class imbalance without altering the original semantic meaning of commit messages. After label normalization, 95,959 instances remain.

Finally, we unify formatting styles by correcting inconsistent whitespace, delimiters, and casing, while stripping redundant appended metadata such as pull-request IDs. All samples are standardized into the canonical CCS structure.

\subsection{Detailed Semantic Filtering Process}
We adopt a two-stage semantic quality control strategy to guarantee semantic matching between code diffs and commit messages.

In the first embedding-based screening stage, we employ Qwen3-0.6B to compute similarity between code diffs and commit subjects. According to the similarity score distribution, we preserve all samples with a similarity score higher than 0.4. All low-score and unscorable samples are forwarded to the second-stage LLM-assisted evaluation.

In the second stage, Qwen2.5-Coder-7B conducts comprehensive quality assessment from four dimensions: accuracy, specificity, conciseness, and completeness, scoring each sample from 0 to 10. A total of 26,185 ambiguous samples are evaluated. Only samples with scores higher than 6 are retained. Finally, we manually inspect a random sample to verify the effectiveness of the semantic filtering process.

%\subsection{Full Final Dataset Statistics}
%After all filtering pipelines, the final dataset contains 86,688 valid high-quality commit-diff pairs. All token statistics are calculated via the Qwen3-0.6B tokenizer.

%In terms of diff length distribution, 57.53\% of diffs contain fewer than 1,024 tokens, while 12.00\% exceed 4,096 tokens, covering both small local edits and large-scale system modifications. For commit subjects, 81.14\% range from 4 to 11 tokens, and 96.24\% are shorter than 16 tokens, which conforms to the concise writing habit of industrial commit messages.

%The commit type distribution presents a natural long-tailed pattern dominated by \texttt{fix}, with relatively fewer \texttt{ci}, \texttt{style}, and \texttt{update} cases. We preserve this real-world distribution instead of artificial balancing. The dataset is finally stratified by commit type and split into training, retrieval, validation, and test subsets following a 6:2:1:1 ratio.

\section{Implementation Setting} \label{app:implementation}
All experiments are conducted on a server with two AMD EPYC 7H12 64-Core CPUs and eight NVIDIA A100 80GB PCIe GPUs, and each experiment is repeated three times. The decoding temperature of all LLM generators is set to 0.8. The easy and hard stages use batch sizes of 96 and 48, 8 and 5 training epochs, learning rates of $2\times10^{-5}$ and $1\times10^{-5}$, and temperatures of 0.05 and 0.03, respectively, with a warmup ratio of 0.06 for both. We set the maximum number of hard negatives $K_i$ in Eq.~(12) to 6, the weighting coefficient $\lambda$ in Eq.~(13) to 1.5, and the random seed to 42. All hyperparameters are selected based on validation performance.

\begin{figure*}[h!]
    \centering
    \begin{minipage}{0.98\textwidth}

    \noindent\textbf{Code Change}
    \begin{lstlisting}[style=casestudydiff]
@@ -64,12 +64,15 @@ describe('Browser Builder errors', () => {
     const run = await architect.scheduleTarget(
         targetSpec, { aot: true }, { logger });
     const output = await run.result;
     expect(output.success).toBe(false);

+    // Wait for the builder to complete
+    await run.stop();
+
     if (!veEnabled) {
       expect(logs.join()).toContain(
           'selector must be a string');
     } else {
       expect(logs.join()).toContain(
           'Function expressions are not supported in');
     }
-    await run.stop();
   });

   it('shows missing export errors', async () => {
@@ -87,7 +90,10 @@ describe('Browser Builder errors', () => {
     const run = await architect.scheduleTarget(
         targetSpec, overrides, { logger });
     const output = await run.result;
     expect(output.success).toBe(false);
-    expect(logs.join()).toContain(
-        `export 'missingExport' was not found in 'rxjs'`);
+
+    // Wait for the builder to complete
     await run.stop();
+
+    expect(logs.join()).toContain(
+        `export 'missingExport' was not found in 'rxjs'`);
   });
 });
    \end{lstlisting}

    \vspace{0.4em}
    \noindent\textbf{Commit Message Comparison}

    \vspace{0.2em}
    \small
    \setlength{\tabcolsep}{4pt}
    \renewcommand{\arraystretch}{1.18}

    \begin{tabularx}{\textwidth}{
        >{\raggedright\arraybackslash}p{0.2\textwidth}
        >{\raggedright\arraybackslash}p{0.08\textwidth}
        >{\raggedright\arraybackslash}p{0.25\textwidth}
        X
    }
        \toprule
        \textbf{Output}
        & \textbf{Type}
        & \textbf{Scope}
        & \textbf{Subject} \\
        \midrule
        \rowcolor{gray!10}
        Ground Truth
        & test
        & @angular-devkit/build-angular
        & ensure jobs are complete before checking logs \\

        Subject-only
        & ---
        & ---
        & ensure builder stops after each test \\

        CCS
        & refactor
        & @angular-devkit/build-angular
        & ensure builder completion before assertions \\

        \mbox{\textsc{CoCoRerank}~Top-1}
        & test
        & @angular-devkit/build-angular
        & ensure builder stops before checking logs \\

        \bottomrule
    \end{tabularx}

    \end{minipage}

    \caption{A case study comparing the ground-truth commit message
    with the outputs generated under different settings.}
    \label{fig:case_study_builder_completion}
\end{figure*}

\section{Case Study}
To qualitatively examine the effects of conventional generation and consistency-based reranking, we conduct a case study on a representative commit. As shown in Figure~\ref{fig:case_study_builder_completion}, we compare the ground-truth commit message with the outputs produced by subject-only generation, joint CCS generation, and \textsc{CoCoRerank}.
\paragraph{Ground-Truth Interpretation.}
The change moves \texttt{await run.stop()} before the log-related assertions, ensuring that the builder has fully completed before its logs are checked. Therefore, the ground-truth subject accurately summarizes the purpose of the modification. The type \texttt{test} is appropriate because the change only adjusts the asynchronous execution order within a test suite, while the scope \texttt{@angular-devkit/build-angular} corresponds to the Browser Builder component being tested.

\paragraph{Progressive Improvement in Subject Quality.}
The Subject-only output, "ensure builder stops after each test" identifies the \texttt{run.stop()} operation but misinterprets its purpose. The change is not simply to stop the builder after each test, but to ensure completion before the logs are checked.

The CCS output, "ensure builder completion before assertions" is more accurate but still overly general, since \texttt{expect(output.success).toBe(false)} remains before \texttt{run.stop()}. It also incorrectly predicts the type as \texttt{refactor}, overlooking the test-specific context.

In contrast, the \textsc{CoCoRerank} output, "ensure builder stops before checking logs" precisely captures both the relocated operation and the affected log assertions.

\paragraph{Effect of Joint CCS Generation.}
Jointly generating the CCS fields provides concrete semantic cues for subject generation. In this case, the scope \texttt{@angular-devkit/build-angular} anchors the subject to the Browser Builder, helping the model retain the specific object builder rather than producing a generic description of test execution. Meanwhile, although \texttt{refactor} is not the most accurate type, it reflects that the change mainly reorders existing statements rather than adding new behavior, encouraging the subject to describe the altered execution order through builder completion before assertions. Consequently, the CCS output captures both the affected component and the temporal relationship more accurately than the Subject-only output.

\paragraph{Effect of Consistency-Based Reranking.}
The CCS output before consistency reranking contains an internal semantic conflict. The word 'assertions' strongly indicates test behavior and is therefore more naturally associated with the type \texttt{test} than with \texttt{refactor}. The generated message describes the execution order of test assertions while simultaneously classifying the change as a refactoring operation. In comparison, the \textsc{CoCoRerank} output forms a coherent combination across all three fields: \texttt{test} corresponds to the log-checking behavior, \texttt{@angular-devkit/build-angular} corresponds to the builder, and "builder stops before checking logs" connects the structural fields through a specific test action.

This advantage remains visible after the initially predicted type \texttt{refactor} is corrected to \texttt{test} through vertical consistency. Although "before assertions" is compatible with the \texttt{test} type, it remains generic because assertions may occur in tests for any component. By contrast, "before checking logs" simultaneously expresses a test-side verification action and a builder-related runtime artifact. The final subject therefore satisfies the joint semantic constraints imposed by both the type and the scope more specifically, illustrating the effectiveness of consistency-based reranking.

\section{Full Prompts}
\label{app:full_prompts}

Here we show the full prompts used in our experiments. 
Instance-specific inputs and retrieved examples are represented by 
placeholders enclosed in angle brackets.

\subsection{LLM-Assisted Quality Assessment}
\label{app:quality_prompt}

The prompt used for LLM-assisted semantic quality assessment is shown below.

\begin{lstlisting}[]
You are a strict data quality evaluator for commit message generation datasets.
Evaluate whether the given commit message is a high-quality label for the given git commit diff.
Scoring rubric:
10 = The message is accurate, specific, concise, and fully matches the code change.
7-9 = Mostly correct, with minor missing details or minor wording issues.
4-6 = Partially related, but vague, incomplete, or misses important parts of the change.
1-3 = Mostly wrong, misleading, too generic, noisy, or barely related to the diff.
0 = Empty, corrupted, unrelated, non-English noise, or impossible to judge as a valid label
You must output exactly one integer from 0 to 10.
Do not output any explanation, punctuation, label, JSON, markdown, or extra text.
Commit Message:
<commit_message>
Commit Diff:
<commit_diff>
Score:
\end{lstlisting}

\subsection{Retrieval-Augmented Candidate Generation}
\label{app:retrieval_prompt}

For retrieval-augmented candidate generation, five retrieved commit--message pairs are inserted into the prompt as in-context demonstrations.

\begin{lstlisting}[]
You are an expert software engineer.Please generate a commit message with the form of "type(scope): subject".
Type is the change type of the commit, which must be chosen from:
{refactor, perf, fix, feat, test, docs, style, chore, build, ci, update}.
Below is the meaning of each type:

- build: changes that affect the build system or external dependencies (example scopes: gulp, broccoli, npm)
- fix: a bug fix
- feat: a new feature
- ci: changes to CI configuration files and scripts (example scopes: Travis, Circle, BrowserStack, SauceLabs)
- perf: a code change that improves performance
- refactor: a code change that neither fixes a bug nor adds a feature
- style: changes that do not affect the meaning of the code (e.g., formatting, whitespace, missing semicolons)
- test: adding missing tests or correcting existing tests
- docs: documentation only changes
- chore: changes to build process, dependencies, or auxiliary tools
- update: general updates or maintenance changes that don't fit other types

Scope is the change scope of the commit.
Subject is a short summary of the code change of the commit.

Below are some examples of git commits and their corresponding commit messages.

[Example 1]
Commit:
<retrieved_commit_1>
Commit Message:
<retrieved_message_1>

[Example 2]
Commit:
<retrieved_commit_2>
Commit Message:
<retrieved_message_2>

[Example 3]
Commit:
<retrieved_commit_3>
Commit Message:
<retrieved_message_3>

[Example 4]
Commit:
<retrieved_commit_4>
Commit Message:
<retrieved_message_4>

[Example 5]
Commit:
<retrieved_commit_5>
Commit Message:
<retrieved_message_5>

Now, given the following commit, generate the commit message.
Your response must not output anything other than the commit message with the form type(scope): subject.

Commit:
<target_commit>
\end{lstlisting}

\subsection{Subject-Only Commit Message Generation}
\label{app:subject_only_prompt}

For the subject-only setting, we use the following prompt without requiring the generated message to follow the CCS structure.

\begin{lstlisting}[]
You are an expert software engineer.Please generate commit message, which is a short summary of the code change of the commit.

Below are some examples of git commits and their corresponding commit messages.

[Example 1]
Commit:
<retrieved_commit_1>
Commit Message:
<retrieved_message_1>

[Example 2]
Commit:
<retrieved_commit_2>
Commit Message:
<retrieved_message_2>

[Example 3]
Commit:
<retrieved_commit_3>
Commit Message:
<retrieved_message_3>

[Example 4]
Commit:
<retrieved_commit_4>
Commit Message:
<retrieved_message_4>

[Example 5]
Commit:
<retrieved_commit_5>
Commit Message:
<retrieved_message_5>

Now, given the following commit, generate the commit message.
Your response must not output anything other than the commit message of the commit.

Commit:
<target_commit>
\end{lstlisting}

\subsection{Hard-Negative Set Generation}
\label{app:hard_negative_prompt}

For hard-negative set generation, we use the same prompt as for retrieval-augmented candidate generation.

% \end{document}

\end{document}